\documentclass[sigconf, nonacm]{acmart}

\newcommand\vldbavailabilityurl{}

\usepackage[disable]{todonotes}
\usepackage{outlines}
\usepackage[htt]{hyphenat}
\usepackage{listings}
\usepackage{url}
\title{xpSHACL: Explainable SHACL Validation using Retrieval-Augmented Generation and Large Language Models}

\author{Gustavo Correa Publio}
\orcid{0000-0002-3853-3588}
\affiliation{%
  \institution{University of Leipzig}
  \state{Germany}
  \postcode{04105}
}
\email{gustavo.publio@informatik.uni-leipzig.de}

\author{José Emilio Labra Gayo}
\orcid{0000-0001-8907-5348}
\affiliation{%
  \institution{University of Oviedo}
  \country{Spain}
}
\email{labra@uniovi.es}

\begin{document}
\begin{abstract}
Shapes Constraint Language (SHACL) is a powerful language for validating RDF data. Given the recent industry attention to Knowledge Graphs (KGs), more users need to validate linked data properly. However, traditional SHACL validation engines often provide terse reports in English that are difficult for non-technical users to interpret and act upon. This paper presents xpSHACL, an explainable SHACL validation system that addresses this issue by combining rule-based justification trees with retrieval-augmented generation (RAG) and large language models (LLMs) to produce detailed, multi-language, human-readable explanations for constraint violations. A key feature of xpSHACL is its usage of a Violation KG to cache and reuse explanations, improving efficiency and consistency.
\end{abstract}

\maketitle



\section{Introduction}
\label{section:introduction}
\subsection{The Proliferation of Knowledge Graphs and the Critical Role of Data Validation}
The increasing prevalence of graph-structured data, particularly in RDF graphs, has established a robust foundation for several applications across diverse domains. These applications range from knowledge representation and the complex process of semantic data integration to providing factual knowledge supplied to machine learning (ML) algorithms. This widespread adoption underscores the vital role of technologies like the Shapes Constraint Language (SHACL) in ensuring the quality and reliability of this ever-growing body of information \cite{okulmus2024shacl}, as data cleaning recommendations significantly improve ML algorithms' performances \cite{mohammed2025}. The semantic model inherent in Knowledge Graphs (KGs), as highlighted by \cite{rajabi2024knowledge}, allows organizing information through hierarchical structures based on entities, their properties, and the relationships that connect them. Large-scale KGs facilitate the integration of heterogeneous information sources, an increasingly essential capability in a data-rich environment. Consequently, the need for effective SHACL validation to maintain the integrity and trustworthiness of these KGs is fundamental to provide clean, curated, and reliable data to users and applications. As the volume of data published as KGs continues to expand, the challenge of guaranteeing its consistency and adherence to predefined schemas, often expressed through SHACL constraints, grows in significance for achieving interoperability and fostering dependable knowledge sharing across different systems and applications \cite{gercke2022supporting}.

\subsection{Limitations of Existing SHACL Validation Engines and the Need for Explainability}
While traditional SHACL validation engines effectively identify constraint violations within RDF data, they often fall short in providing users with the necessary understanding of \textit{why} these violations occurred, leading to the need for debugging to understand unexpected violations \cite{rdf4jshacl}. The primary output of these engines, the validation report, typically enumerates what constraints are violated by which data nodes, but often lacks a deep, traceable explanation of why the violation occurred in terms of the interaction between data and SHACL rules. Recent academic efforts have begun to address the interpretability of SHACL outcomes, for instance, by developing formal provenance semantics to explain \textit{why} a given node conforms to a shape \cite{delva2023data}. Such approaches provide valuable insights into the data supporting valid states. However, the challenge of providing intuitive, actionable explanations specifically for violations, especially for users not deeply versed in SHACL semantics and RDF graph structures, remains a significant hurdle. The reports generated by existing engines are frequently terse and filled with technical details. The absence of intuitive explanations creates a significant barrier to the wider adoption and effective use of SHACL by a broader community of data managers and users who require a more accessible and understandable validation process. This limitation hinders the effective utilization of SHACL for crucial tasks such as data curation and the debugging of semantic data, especially for users whose primary expertise lies outside the realm of semantic web technologies, as is frequently the case for domain experts \cite{denaux2011supporting}. 

\begin{figure*}[hpbt!]
  \centering
\includegraphics[width=\textwidth]{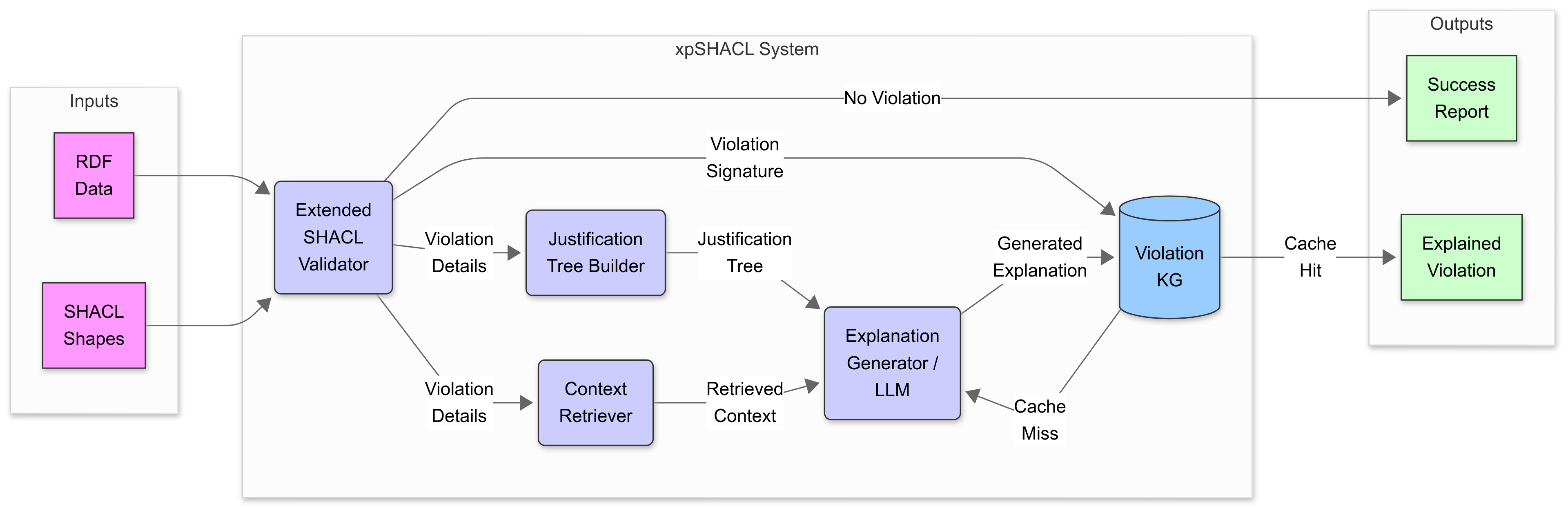}
  \caption{xpSHACL process overview.}
  \label{fig:arch}
\end{figure*}

\subsection{Introducing xpSHACL: An Explainable SHACL Validation System}
To address those limitations, this paper introduces xpSHACL\footnote{Available at \href{https://www.github.com/gcpdev/xpshacl}{https://www.github.com/gcpdev/xpshacl}}, an open-source innovative system designed to provide understandable explanations for SHACL constraint violations. xpSHACL achieves this by employing a novel combination of techniques: 

\begin{itemize}
    \item Justification Tree: Provides the verifiable, logical backbone/trace of why the violation occurred according to SHACL rules. Ensures factual grounding based on the validator's logic.

    \item Retrieval augmented generation (RAG): Enriches this logical trace with contextual domain rules, shape documentation, ontology fragments, and similar past cases, making the explanation more informative and relevant. The KG also adds efficiency and consistency by caching violations and explanations.

    \item Large Language Model (LLM): Translates the structured tree and retrieved context into fluent, human-readable natural (multi) language, potentially adding helpful suggestions or rephrasing complex logic in a simpler way.
\end{itemize}

This synergistic approach aims to bridge the gap between the technical output of SHACL validation and the comprehension needs of a diverse range of users by generating detailed, human-readable explanations that might avoid excessive technical jargon, enabling non-technical users to take action over violations in data. A key feature of xpSHACL is its utilization of a Violation KG, which serves as a persistent repository for caching and reusing multi-language explanations.  This mechanism improves efficiency by avoiding redundant explanation generation and contributes to the consistency of the explanations for recurring violations. Furthermore, a KG is chosen over a traditional cache database for caching LLM explanations because its inherent ability to store and query semantic relationships and contextual information naturally aligns with the interconnected nature of SHACL violations and their complex justifications, enabling more intelligent and contextually relevant retrieval than a simple key-value or tabular store. 

The combination of structured reasoning through justification trees, the enrichment of context via RAG over a KG as a knowledge base with explicit domain rules, and the power of LLMs to articulate explanations in natural language of any wide-spoken language of the world represents a significant advancement in making SHACL validation more accessible and actionable for a broader audience. 

Figure \ref{fig:arch} pictures an overview of the proposed process workflow, while each architectural component is described in detail in Section \ref{section:arch}.

\subsection{Research Problem and Key Contributions}
Given that existing SHACL validation engines often produce reports that are difficult for non-technical users to interpret and act upon (Section 1.2), this paper addresses the central research question:

How can a system combining rule-based justification, retrieval-augmented generation, and LLMs bridge the gap between technical SHACL validation outputs and the comprehension needs of diverse users, enabling the generation of understandable and actionable explanations for constraint violations in RDF data?

To answer this question, this paper makes the following key contributions:

\begin{itemize}
    \item presents the architecture and implementation of xpSHACL, a novel system for explainable SHACL validation
    \item introduces a unique approach that combines rule-based justification trees with retrieval-augmented generation (RAG) over KG and LLMs to generate human-readable explanations for SHACL constraint violations, including support for generating explanations and suggestions in multiple languages.
    \item proposes the usage of a Violation KG as a mechanism for efficiently storing violations, retrieving similar cases and domain rules, and reusing previously generated explanations, thereby improving the performance and consistency of the system.
    \item outlines a comprehensive evaluation plan to assess the effectiveness of xpSHACL in terms of explanation quality, efficiency, consistency, and user satisfaction. 
\end{itemize}

\subsection{Structure of the Paper}
The remainder of this paper is structured as follows: Section \ref{section:related} provides a detailed overview of the related work in the areas of SHACL and RDF validation, explainable AI in semantic web technologies and KGs, retrieval-augmented generation, natural language generation, existing approaches to explainable SHACL validation, and KG enhanced language models. Section \ref{section:arch} describes the architecture of the proposed xpSHACL system and its key components. Section \ref{section:implementation} elaborates on the implementation details of xpSHACL. Section \ref{section:evaluation} presents a comprehensive evaluation plan designed to assess the system's performance and effectiveness. Finally, Section \ref{section:conclusion} concludes the paper and outlines potential directions for future work.




\section{Background}\label{section:background}

This section provides the foundational concepts and established technologies essential for understanding the xpSHACL system. Such background sets the stage by detailing the core building blocks and underlying theories that xpSHACL leverages. 

\subsection{Foundations of SHACL and RDF Validation}
The Shapes Constraint Language (SHACL) is a W3C Recommendation that serves as a declarative language for validating RDF graphs against a defined set of conditions, expressed as shapes. These shapes, which form a shapes graph, validate target nodes within a data graph, ensuring that the data adheres to the specified constraints. The SHACL Core language defines two primary types of shapes: node shapes, which impose constraints on the focus node itself, and property shapes, which define constraints on the values of specific properties or paths originating from the focus node. The W3C specification \cite{w3cshacl} details the terminology, concepts, and language features of SHACL, providing a normative foundation for its use in ensuring RDF data quality. Study indicates that while SHACL has gained significant traction in the industry for validating RDF data, the development of algorithms for SHACL constraint validation is an ongoing area of research \cite{corman2019validating}. 
The semantics of graph validation against SHACL shapes, including the complexities introduced by recursive shapes, have been explored in the literature, providing a clear understanding of what constitutes a valid RDF graph according to a given set of shapes. The standard outcome of the SHACL validation process is a validation report, which details any violations found in the data graph related to the conditions defined in the shapes graph \cite{pareti2021review, gayo2017validating}. 
However, as noted earlier, these reports often lack detailed explanations aimed at non-technical, business-oriented users.

SHACL is designed for data validation and quality assurance, serving a distinct role in the Semantic Web stack compared to languages like OWL and RDFS, primarily focused on defining ontologies and enabling inferencing \cite{polleres2013rdfs}. It allows for the expression of a wide range of constraints, including those related to the existence of properties, the types and ranges of their values, and the cardinality of relationships, in the form of shapes. Various tools and libraries have been developed to support shape creation and validation, such as RML2SHACL \cite{delva2021rml2shacl}, which facilitates the generation of SHACL shapes from RML mappings, and \texttt{pyshacl} \cite{sommer2021pyshacl}, a Python library for validating RDF graphs against SHACL shapes. Furthermore, SHACL validation engines, such as the one implemented in RDF4J \cite{rdf4jshacl}, or RDFUnit \cite{kontokostas2014test}, employ techniques like validation plans and transaction management to perform the validation process efficiently \cite{rdf4jshacl}. These engines analyze changes in data and shapes to optimize validation. Still, their primary focus has traditionally been on the efficiency and correctness of the validation process rather than on providing user-friendly explanations for the resulting violations.

\subsection{Explainable AI in Semantic Web Technologies and Knowledge Graphs}
The field of Explainable AI (XAI) has seen increasing interest in the context of Semantic Web technologies and KGs as a way of closing the gap to neural-symbolic integration with AI models \cite{futia2020integration}. With their inherent structure and semantic richness, KGs are recognized as a powerful means of facilitating access to and integration of data, adding crucial context to other AI techniques, and generating understandable explanations to humans. Research has highlighted the potential of KGs in explainable AI systems due to their ability to represent knowledge through hierarchical structures and to integrate information from diverse and heterogeneous sources \cite{rajabi2024knowledge,futia2020integration}. The structured nature of KGs allows encoding relationships and semantic context, making them natively more explainable compared to black-box models like deep learning networks. RDF and other Semantic Web technologies offer a strong foundation for building explainable AI systems where transparency in decision-making processes is fundamental by providing formal semantics for representing and reasoning about data \cite{ebrahimi2018reasoning}. KGs serve as a cornerstone of explainable AI by providing a simple yet homogeneous data model that can be systematically exploited to understand the reasoning behind AI systems. A comprehensive review of knowledge-graph-based explainable AI has categorized its applications into areas such as feature extraction, relationship extraction, KG construction, and KG reasoning, demonstrating the broad utility of this approach across different stages of AI model development and deployment \cite{rajabi2024knowledge}. The growing academic and industrial interest in this intersection is further evidenced by dedicated books and research focused on the foundations, applications, and challenges of using KGs for explainable AI \cite{tiddi2020knowledge,rajabi2024knowledge,bianchi2020knowledge}. These efforts aim to leverage the semantic understanding provided by KGs to generate results from AI systems that are accurate, interpretable, and trustworthy for users in various domains.

\subsection{Retrieval-Augmented Generation (RAG) for Knowledge-Intensive Tasks}
Retrieval-Augmented Generation (RAG) has emerged as a significant technique for enhancing the capabilities of language models, particularly in knowledge-intensive natural language processing tasks \cite{zhu2024structugraphrag}. This approach involves grounding the language model's generation process in external knowledge retrieved from a knowledge base or a set of documents, allowing it to produce more accurate, contextually relevant, and informative responses. \cite{chen2024optimizing} shows that RAG can be particularly powerful in the context of KGs. By retrieving relevant subgraphs or information from the KG based on a user's query or the task at hand, the language model can generate responses rooted in factual knowledge and leverage the relationships and semantic context encoded within the graph. Research has shown that integrating KGs into RAG systems can address some of the limitations of traditional document-based RAG, such as the loss of original content context and the inability to effectively capture relationships between entities that do not co-occur in the same document. Techniques such as node and relation extraction, graph clustering, and summarization are used to retrieve relevant information from KGs for augmentation. Using KGs in RAG also enables deductive reasoning capabilities, allowing the AI to make decisions based on established rules and relationships within the graph, potentially leading to more verifiable and contextually rich explanations \cite{chen2024optimizing}. The growing interest in this area is reflected in surveys and tech reports that provide overviews of methodologies for constructing KGs for RAG applications and the steps involved in their implementation \cite{chen2025retrieval,matsumoto2024kragen,bratanic2024using}. Furthermore, tools like KRAGEN demonstrate the potential of combining KGs, RAG, and advanced prompting techniques to solve complex problems by leveraging structured knowledge for more accurate and explainable outcomes \cite{matsumoto2024kragen}.

\subsection{Natural Language Generation (NLG) for Explaining Semantic Data}
Natural Language Generation (NLG) is crucial in transforming structured data and formal representations into human-readable text, making complex information more accessible to a wider audience \cite{gercke2022supporting}. In the context of SHACL validation, NLG techniques, particularly those powered by LLMs, can be instrumental in explaining the meaning and implications of SHACL constraints and the reasons behind any violations that occur. While some research focuses on the reverse process of automatically generating SHACL shapes from natural language descriptions of data requirements, the ability to generate natural language explanations from SHACL constraints and validation results is essential for improving the usability of SHACL for non-technical users \cite{donkers2024converting}. The SHACL specification itself provides examples of how SHACL constraints can be described in natural language through \texttt{sh:name} and \texttt{sh:description}, indicating an inherent connection between the formal language and human-readable descriptions \cite{w3cshacl}. The application of SHACL constraints in analyzing the quality of generated ontologies also suggests a role for NLG in reporting these quality assessments in a comprehensible manner \cite{dasilva2024use}. Thus, the ability to translate the formal semantics of SHACL and its validation outcomes into natural language is a key component in making SHACL more accessible and actionable for users who may not have a deep technical understanding of semantic web technologies.

\subsection{Explanations generation from Rule-Based Systems}
Generating explanations from rule-based systems is a well-studied area in artificial intelligence. Rule-based systems operate on predefined rules to make decisions or draw conclusions and are often favored for their inherent transparency, understandability, and explainability \cite{xaqt2023mastering}. In these systems, explanations can be provided by tracing the rules applied to reach a particular outcome. Research has explored various frameworks and techniques for generating explanations from rule-based systems, including topological frameworks that characterize explainability in terms of the definability of a classifier relative to an explanation scheme \cite{mullins2023shape}. Another area of focus is generating contrastive explanations, which aim to answer why a particular outcome occurred rather than an expected alternative \cite{herbold2024generating}. These types of explanations can be particularly useful in scenarios where users have specific expectations about the system's behavior.

As discussed by \cite{mullins2023shape}, the fundamental structure of a rule-based system, typically consisting of a set of "if-then" rules and an inference engine that applies these rules to a knowledge base, lends itself to providing explanations by revealing the chain of reasoning. This broader body of research on explainable rule-based systems is relevant to xpSHACL's use of justification trees, which essentially represent the rule-based logic behind SHACL constraint violations in a structured format that can then be used to generate natural language explanations.

\subsection{KG Enhanced LLMs}
The synergy between KGs and LLMs has emerged as a powerful paradigm for enhancing the performance and interpretability of NLP systems. LLMs can be grounded in factual information by leveraging the structured knowledge encoded in KGs, leading to more faithful and accurate reasoning \cite{luo2023reasoning}. Frameworks like KELDaR propose using KGs to provide explainable reasoning for LLMs in tasks such as KG question answering. These approaches often involve retrieving relevant subgraphs from the KG to assist the LLM in answering complex questions, thereby explaining the knowledge used in the reasoning process \cite{li2024framework}. Similarly, methods like GLAME explore how KGs can be used to enhance the ability of LLMs to incorporate new knowledge through model editing, ensuring that the changes are consistent with the structured relationships in the graph \cite{zhang2024knowledge}. The overall trend in this area is to take advantage of the vast amounts of knowledge captured in KGs to improve the reasoning capabilities of LLMs and to make their decision-making processes more transparent and understandable.


\section{Related Work}\label{section:related}

We focus on existing research and systems that have attempted to provide explainable SHACL validation, highlighting their approaches and limitations as direct predecessors or alternatives to xpSHACL.

Research has specifically addressed the challenge of explaining SHACL validation outcomes, particularly in cases of non-validation \cite{ahmetaj2021reasoning}. One prominent approach involves explaining non-validation through the concept of repairs, where an explanation is formulated as a minimal set of additions and deletions of facts to the original RDF graph that would result in its validation against the given SHACL constraints \cite{ahmetaj2021reasoning,ahmetaj2024consistent}.

Theoretical work has also analyzed the computational complexity of reasoning about these repairs in the context of SHACL constraints \cite{ahmetaj2024consistent}. Another related approach to explain RDF validations involves using rule-based reasoning systems \cite{demeester2021rdf}. These systems, such as those based on N3Logic and the EYE reasoner, aim to generate logical proofs that trace the reasoning steps leading to a constraint violation \cite{berners2008n3logic,demeester2021rdf}. By providing a detailed logical derivation, these systems offer a transparent explanation of \textit{why} a particular violation occurred, often down to the specific parts of the RDF graph and the axioms of the ontology that triggered the inference leading to the violation \cite{demeester2021rdf}. While these existing approaches, based on logic programming and rule-based reasoning, offer formal and computationally grounded methods for explaining SHACL violations, they often result in explanations in a formal language or require familiarity with logical concepts to be comprehended.

In contrast, xpSHACL aims to explain validations in natural language using retrieval-augmented generation and LLMs, offering a potentially more accessible alternative for a broader range of users.

\section{xpSHACL Architecture}\label{section:arch}
The xpSHACL architecture is designed to provide detailed and understandable explanations for violations of SHACL constraints in RDF data. It achieves this through the interaction of several key components, each playing a specific role in the explanation generation process.

\todo[inline]{4.3 Context Enrichment:

Line 535 (rdflib queries): Good to show examples.

Line 548 (Similar Case Retrieval): How is "similarity" defined? Same node type + same violation type? Needs clarification.

Line 575 (xsh:appliesToProperty): Clarify this property/prefix. Is it assumed to exist in the shapes graph?

How are the different retrieved pieces of context (ontology, docs, similar cases, rules) combined or prioritized before being fed to the LLM? Is it simple concatenation, or is there a more structured prompting strategy?}
\subsection{Extended SHACL Validator}
The process begins with an Extended SHACL Validator, which builds upon the functionality of a standard SHACL validation engine. While a typical validator would identify whether an RDF data graph conforms to a given shapes graph and report any violations, the Extended SHACL Validator is enhanced to capture a more granular level of detail about each violation. This includes the fact that a violation occurred and crucial information such as the specific shape and violated constraint, the particular focus node and property path involved in the violation, and the actual value in the data that triggered the failure. This detailed context is essential as it forms the foundation on which the subsequent explanation generation process is built. By providing precise information about the nature and location of the violation, this extended validator ensures that the system has the necessary input to construct accurate and relevant explanations.

\subsection{Justification Tree Builder}
Once a violation is detected and its details are captured by the Extended SHACL Validator, the information is passed to the Justification Tree Builder. This component is responsible for constructing a logical justification tree for each violation. The justification tree represents a step-by-step breakdown of the reasoning process, based on the relevant SHACL rules and the structure of the RDF data, that led to the identification of the violation. Each node within the tree corresponds to either a logical inference made during the validation process or the application of a specific SHACL constraint. The tree root represents the final violation, and the branches trace back the conditions and rules evaluated to arrive at this conclusion. This structured, rule-based representation of the violation's derivation provides a transparent view of the underlying logic, effectively explaining \textit{why} the violation occurred according to the defined SHACL shapes and the data being validated.

\subsection{Retrieval-Augmented Generation for Context Enrichment}

xpSHACL leverages Retrieval-Augmented Generation (RAG) to enrich the context provided to the LLM when generating explanations. This process involves retrieving relevant information from several sources to provide a comprehensive understanding of the violation.

\subsubsection{Ontology Fragment Retrieval}

Relevant triples from the data graph that involve the focus node of the violation are retrieved. This provides the LLM with information on the properties and relationships of the entity involved in the violation. For example, for a violation related to \texttt{ex:resource1}, triples such as 
\footnote{We use Turtle notation and omit prefix declarations for clarity.}
\begin{lstlisting}[breaklines]
ex:resource1 a ex:Person .
\end{lstlisting} 
might be retrieved to provide context about the entity's type. This is achieved using \texttt{rdflib} queries like:
\begin{lstlisting}[breaklines]
SELECT ?p ?o WHERE {
  ex:resource1 ?p ?o . 
}
\end{lstlisting}

\subsubsection{Shape Documentation Retrieval}

If the SHACL shape associated with the violation has documentation (e.g., using \texttt{rdfs:comment}), this is retrieved to provide additional information about the purpose and constraints defined by the shape.

\subsubsection{Similar Case Retrieval}\label{section:similar-case-retrieval}

The system identifies other data instances similar to the focus node that exhibit the same violation pattern, i.e., the core characteristics of a violation, specifically the type of constraint violated, the associated property path, and the general violation type (e.g., Cardinality), abstracted from specific data instances. This helps the LLM understand the broader context of the violation and identify potential patterns or common issues in the data. For example, if a violation occurs because a \texttt{Person} is missing the \texttt{ex:hasName} property, the system retrieves other \texttt{Person} instances that also lack this property. This is done using SPARQL queries such as:

\begin{lstlisting}[breaklines]
SELECT DISTINCT ?node WHERE {
  ?node a ex:Person .
  FILTER NOT EXISTS { 
    ?node ex:hasname ?anyval . 
  }
  FILTER(?node != ex:resource1)
}
\end{lstlisting}

\subsubsection{Domain Rule Retrieval}\label{section:domainrule}

Domain-specific rules that are relevant to the violated property or constraint are retrieved from the shapes graph. These rules provide further context about the constraints and any domain-specific knowledge that might be relevant to the violation. For instance, if there's a rule stating that "the \texttt{ex:hasAge} property must be a non-negative integer," this rule is retrieved to provide additional context for a violation related to age. The system uses queries like:

\begin{lstlisting}[breaklines]
SELECT DISTINCT ?rule ?comment
WHERE {
 ?rule xsh:appliesToProperty ex:hasAge .
 OPTIONAL {
  ?rule rdfs:comment ?comment . 
 }
}
\end{lstlisting}

\subsection{Impact of Retrieved Context on Explanation Generation}
The aforementioned contextual elements retrieved by the \textit{Context Retriever} play a crucial synergistic role in enhancing the quality and relevance of the explanations generated by the LLM. These elements provide the LLM with a richer understanding of the violation beyond the mere structural details of the SHACL report and justification tree.

\begin{itemize}
    \item \textbf{Ontology fragments} about the focus node ground the LLM's understanding in the specific data instance. By providing direct triples related to the violated entity, the LLM can generate explanations that are factually aligned with the current state of the KG, detailing what specific data led to the violation without exposing more data than necessary, addressing privacy concerns about sharing or leaking proprietary data to external endpoints.
    \item \textbf{Shape documentation} (e.g., \texttt{rdfs:comment}, \texttt{sh:name}) offers the human-intended purpose or descriptive information about the SHACL shapes themselves. This allows the LLM to infuse the explanation with the 'why' behind a constraint, making it more intuitive and less purely technical for users.
    \item \textbf{Similar cases} provide the LLM with a broader perspective on recurring violation patterns within the dataset. Associating violation types enables the LLM to generate more generalized explanations and correction suggestions that address common data quality issues, thereby improving the actionability of the advice.
    \item \textbf{Domain rules}, explicitly defined in the shapes graph, provide the LLM with access to higher-level business logic or best practices. This allows explanations to go beyond the mechanics of SHACL, incorporating domain-specific rationale that is highly valuable for domain experts.

\end{itemize}

By integrating these diverse information contexts, the LLM can synthesize explanations that are not only accurate (grounded by the justification tree) but also contextually rich, human-readable, and directly actionable, fulfilling xpSHACL's goal of bridging the gap between technical validation outputs and user comprehension.

\subsection{Explanation Generator}
The core of the explanation generation process lies within the Explanation Generator component, which utilizes LLMs. This component takes as input the detailed information about the violation from the Extended SHACL Validator, the structured reasoning provided by the Justification Tree Builder, and the relevant domain knowledge retrieved by the Context Retriever. The LLM is then prompted to synthesize this information into a natural language explanation that is designed to be clear, concise, and easily understandable by users, regardless of their technical expertise in SHACL or semantic web technologies.

In addition to explaining the reason for the violation, the Explanation Generator can also leverage the retrieved context and the logical structure of the justification tree to suggest potential corrections or actions that the user can take to resolve the violation and ensure data compliance. The use of LLMs allows for the generation of explanations that are not only accurate but also tailored to human comprehension, bridging the gap between the formal technical details of a SHACL violation and the practical understanding required by data curators and users.

Finally, it is designed to handle requests and generate output in multiple specified languages of both explanations and suggestions.

\todo[inline]{Clarity of the "Violation Knowledge Graph" (KG):

This is a key component, but its exact structure and content need more definition earlier (perhaps in the Architecture section).

What is a "Violation Signature"? (Lines 614, 620, 656, 800). This is critical for caching. Is it a hash? A tuple of (Shape IRI, Constraint Component IRI, Focus Node Type, Property Path)? Be precise. How robust is it to minor variations?

What exactly is stored? Lines 611-613 state it stores "information about previously encountered... violations, besides domain rules". Line 631 says the "generated explanation, paired with its signature, is persisted". Does it store:

The signature? (Yes)

The generated natural language explanation? (Yes, per line 631)

The justification tree? (Maybe? Useful for consistency checks or future interactive exploration).

The retrieved context used for generation? (Maybe? For reproducibility/debugging).

The "domain rules" mentioned in 4.3.4 and 613? Are these copied from shapes or defined separately? Clarify the KG's scope and schema.

The xsh: prefix: (Line 575) Used in a SPARQL query example. Define this prefix. Is it the namespace for the Violation KG's ontology?}

\subsection{Violation Knowledge Graph}
The Violation KG is a crucial component for improving the efficiency and consistency of xpSHACL. It is a persistent knowledge base that stores information about previously encountered SHACL constraint violations and domain rules as described in Section \ref{section:domainrule}. The KG is structured to store violation signatures, which are based on key characteristics of the violation, such as the specific shape and constraint that were violated, the property path involved, and potentially the type of data that caused the failure. When a new violation occurs, xpSHACL first generates a signature for it and then queries the Violation KG to check if an explanation for a similar violation already exists. If a match is found, the cached explanation, along with any suggested corrections, is retrieved from the KG and presented to the user, significantly reducing the time required to generate an explanation. In the absence of a pre-existing explanation matching the detected violation signature, xpSHACL initiates a process involving the Justification Tree Builder, Context Retriever, and Explanation Generator. Subsequently, the generated explanation, paired with its signature, is persisted within the Violation KG for subsequent retrieval and application, with its related language tag (each explanation-suggestion pair may have multiple entries, but no more than one for each language). This caching and reuse mechanism not only enhances the performance of xpSHACL but also ensures a greater degree of consistency in how similar violations are explained over time, which is substantially important given the non-deterministic characteristic of LLMs answers \cite{song2024good}, contributing to a more predictable and reliable user experience.

\subsection{Data Flow}
The process within xpSHACL begins when RDF data and SHACL shapes are provided as input to the Extended SHACL Validator. This component performs the validation and, upon detecting any violations, extracts detailed information about them. For each identified violation, the extracted details are passed to the Justification Tree Builder, which constructs a logical justification tree representing the reasoning behind the failure. Simultaneously, the Context Retriever identifies and retrieves relevant domain knowledge about the violation from sources such as ontologies and shape documentation. Once this information is gathered, xpSHACL generates a unique signature for the violation based on its key characteristics. The system then queries the Violation KG using this signature to determine if an explanation for a similar violation has been previously generated and stored. If a matching explanation is found in the KG, it is directly retrieved. If not, the detailed violation information, the justification tree, and the retrieved context are fed into the Explanation Generator, which utilizes an LLM to produce a natural language explanation and suggest potential corrections. This newly generated explanation is stored in the Violation KG, associated with the violation's signature, for future use. Finally, regardless of whether the explanation was retrieved from the KG or newly generated, xpSHACL outputs this detailed explanation to the user, providing them with a clear understanding of the SHACL constraint violation and guidance on how to address it.

\section{Implementation}
\label{section:implementation}

This section outlines the xpSHACL implementation, built on the architecture described in Section \ref{section:arch}. Implemented in Python 3, it uses established libraries for RDF processing, SHACL validation, and LLM interaction. The open-source code is at \url{https://www.github.com/gcpdev/xpshacl}, as noted in Section \ref{section:introduction}.

\subsection{Core Libraries}
Key Python libraries include:
\begin{itemize}
    \item \textbf{rdflib}: For RDF graph parsing, manipulation, and querying (data, shapes, Violation KG), and SPARQL execution.
    \item \textbf{pyshacl}: The underlying engine for standard SHACL validation.
    \item \textbf{openai}: Client for LLMs supporting the OpenAI API (e.g., ChatGPT, Gemini, Claude) used by \texttt{ExplanationGenerator}.
    \item \textbf{ollama}: Client for locally hosted LLMs via the Ollama framework, used by \texttt{LocalExplanationGenerator}.
    \item \textbf{python-dotenv}: Manages environment variables, especially API keys.
    \item \textbf{Faker}: For generating synthetic test data.
\end{itemize}

\subsection{Extended SHACL Validator (\texttt{ExtendedShaclValidator})}
This component (\path{src/extended_shacl_validator.py}) wraps \\ \texttt{pyshacl.validate}, parses its report graph, and extracts detailed information from \path{sh:ValidationResult} nodes (e.g., \path{sh:focusNode}, \path{sh:sourceShape}, \path{sh:value}). It categorizes a \texttt{ViolationType} based on \texttt{sh:sourceConstraintComponent} and encapsulates this, along with context like constraint parameters, into a \texttt{ConstraintViolation} data class.

\subsection{Justification Tree Builder (\texttt{JustificationTreeBuilder})}
Located in \path{src/justification_tree_builder.py}, this takes a \texttt{ConstraintViolation} and constructs a logical trace of the violation. Using \texttt{rdflib} to query data and shapes graphs, it builds a \texttt{JustificationTree} with nodes representing:
\begin{itemize}
    \item \textbf{Premises}: From SHACL shape definitions.
    \item \textbf{Observations}: Facts from the data graph with evidence.
    \item \textbf{Inferences}: Logical steps connecting premises and observations.
\end{itemize}

\subsection{Context Retriever (\texttt{ContextRetriever})}
This component (\path{src/context_retriever.py}) gathers additional information around a violation via SPARQL queries to make LLM explanations more informative. It retrieves ontology fragments related to the focus node, \path{rdfs:comment} documentation from shapes, similar violation cases (as per Section \ref{section:similar-case-retrieval}), and human-readable domain rules (linked via \path{xsh:appliesToProperty}). This is stored in the \texttt{DomainContext} data class. The retrieval of ontology fragments is not only important for context enhancement, but also to avoid exposing more data than needed to LLMs due to privacy concerns.

\subsection{Explanation Generator (\texttt{ExplanationGenerator} / \texttt{LocalExplanationGenerator})}
These classes (\path{src/explanation_generator.py}) interface with LLMs (via \texttt{openai} or \texttt{ollama}) to convert the \texttt{ConstraintViolation}, \texttt{JustificationTree}, and \texttt{DomainContext} into natural language.
Structured prompts guide LLMs to explain \textit{why} a violation occurred and suggest \textit{how} to fix it, generalizing from specific instances. The output is structured using the \texttt{ExplanationOutput} data class with multilingual explanations, suggestions, and optional traceability data, noting the \texttt{provided\_by\_model}.

\subsection{Violation KG (\texttt{ViolationKnowledgeGraph})}
\label{section:impl-vkg}
Implemented in \path{src/violation_kg.py}, this \texttt{rdflib}-based caching mechanism uses an RDF graph (\path{data/validation_kg.ttl}) and an ontology (\path{data/xpshacl_ontology.ttl}) defining classes like \path{xsh:ViolationSignature} and \path{xsh:Explanation}.
A unique \texttt{ViolationSignature} URI (MD5 hash of constraint component, property path, violation type) identifies a violation type.
\begin{itemize}
    \item \textbf{Storage}: Newly generated explanations, along with their signature components, \texttt{naturalLanguageText}, \texttt{correctionSuggestions}, \texttt{providedByModel}, and serialized input data (violation, tree, context as JSON), are stored under the signature URI in the KG.
    \item \textbf{Retrieval}: Before generation, the KG is checked using the signature. If found, the stored explanation and suggestions are retrieved to reconstruct an \texttt{ExplanationOutput}.
\end{itemize}

\subsection{Orchestration (\texttt{main.py})}
The main script (\path{src/main.py}) handles command-line arguments and coordinates the workflow:
For each \texttt{ConstraintViolation} from the \texttt{ExtendedShaclValidator}:
\begin{enumerate}
    \item Build \texttt{JustificationTree}.
    \item Retrieve \texttt{DomainContext}.
    \item Generate \texttt{ViolationSignature}.
    \item Check \texttt{ViolationKnowledgeGraph}: if cached, retrieve; otherwise, generate new explanation via \texttt{ExplanationGenerator} and store it.
    \item Collect \texttt{ExplanationOutput}.
\end{enumerate}
Final explanations are contained inside the JSON output.

\subsection{Data Structures and Serialization}
Data classes in \path{src/xpshacl_architecture.py} (e.g., \texttt{ConstraintViolation}, \texttt{JustificationTree}, \texttt{ExplanationOutput}) ensure consistent information flow. They use Python dataclasses and include \texttt{to\_dict} methods for serialization (e.g., to JSON for the Violation KG), with current KG retrieval focusing on text components.

\section{Methodology} 
\label{sec:methodology}

The core methodology of xpSHACL centers around augmenting standard SHACL validation with several components designed to produce rich, understandable, and reusable explanations for constraint violations. The system's architecture (visualized in Figure~\ref{fig:arch}) integrates symbolic reasoning (justification trees) with LLMs and retrieval-augmented generation (RAG) principles.

\paragraph{Extended SHACL Validation}
Instead of relying solely on the standard validation report, xpSHACL employs an extended validation layer (implemented in \texttt{src/extended\_shacl\_validator.py}) built upon the \texttt{pyshacl} library \cite{sommer2021pyshacl}. This layer captures detailed information for each violation result provided by \texttt{pyshacl}, including the focus node, source shape, specific constraint component (e.g., \texttt{sh:MinCountConstraintComponent}), result path, violating value, and severity. This structured data forms the input for the subsequent explanation process. We configure the underlying validator to operate with no inferencing (\texttt{inference='none'}) for baseline performance evaluation, although other inference modes supported by \texttt{pyshacl} (\texttt{rdfs}, \texttt{owlrl}) can be used.

\paragraph{Justification Tree Construction}
For each violation identified, a \texttt{JustificationTreeBuilder} component (see \texttt{src/justification\_tree\_builder.py}) constructs a logical justification tree. This tree explicitly represents the reasoning steps leading to the violation conclusion. It typically includes:
\begin{itemize}
    \item A root node stating the main conclusion (e.g., node fails shape conformance).
    \item Premise nodes representing SHACL constraints derived from the shapes graph (e.g., "Shape X requires property P with minCount 1").
    \item Observation nodes representing relevant facts extracted from the data graph (e.g., "Node N has 0 values for property P").
    \item Inference nodes connecting premises and observations to the conclusion (e.g., "Since 0 < 1, the minCount constraint is violated").
\end{itemize}
This symbolic representation provides a traceable and verifiable foundation for the explanation.

\paragraph{Violation Signature and KG}
The Violation KG is managed by the \texttt{ViolationKnowledgeGraph} class (\path{src/violation_kg.py}). Central to this is the concept of a \texttt{ViolationSignature} (defined in \texttt{src/violation\_signature.py}). A signature abstracts the core characteristics of a violation (constraint component type, property path, violation type) independently of the specific data instance (focus node and concrete value) or shape ID.

When a violation occurs, its signature is generated (using \texttt{src/violation\_signature\_factory.py}). The Violation KG (persisted in \texttt{data/validation\_kg.ttl} using the schema from \texttt{data/xpshacl\_ontology.ttl}) is queried for this signature.
\begin{itemize}
    \item \textbf{Cache Hit:} If the signature exists and has a stored explanation in the requested language, the cached natural language explanation and correction suggestions are retrieved directly, significantly speeding up the process
    \item \textbf{Cache Miss:} If the signature is new or lacks an explanation in the target language, the system proceeds to generate a new explanation using the LLM.
\end{itemize}
Newly generated explanations are then stored in the KG, associated with the violation signature and language tag, for future reuse.

\paragraph{Context Retrieval (RAG)}
To enrich the explanations with relevant domain information, the \texttt{ContextRetriever} component implements the RAG aspect. Given a violation, it retrieves contextual snippets, including:
\begin{itemize}
    \item \textbf{Ontology Fragments:} Triples from the data graph related to the focus node.
    \item \textbf{Shape Documentation:} Comments (\texttt{rdfs:comment}) or labels (\texttt{sh:name}) associated with the violated SHACL shape.
    \item \textbf{Similar Cases:} [\textit{Optional: Mention if this feature is actively used/evaluated}] Identifies other nodes in the data graph potentially exhibiting similar patterns (e.g., other nodes of the same type missing the required property).
    \item \textbf{Domain Rules:} [\textit{Optional: Mention if domain rules are defined/used in evaluation}] Looks up relevant domain-specific rules linked to the property path or constraint in the shapes graph or a dedicated rule base.
\end{itemize}
This retrieved context is passed to the LLM alongside the justification tree.

\paragraph{Natural Language Explanation Generation (LLM)}
The final step involves generating human-readable text using an LLM via the \texttt{ExplanationGenerator} or \texttt{LocalExplanationGenerator} classes (\texttt{src/explanation\_generator.py}). The system constructs prompts incorporating the violation details, the justification tree structure, and the retrieved domain context. Specific instruction templates (like \texttt{explanations\_prompt} and \texttt{suggestions\_prompt}) guide the LLM to produce:
\begin{itemize}
    \item A natural language explanation of why the violation occurred, focusing on the violation type rather than specific data values for better generalization.
    \item Actionable correction suggestions tailored to the violation type.
\end{itemize}
The system supports multiple LLM backends via API (OpenAI, Google Gemini, Anthropic) and local execution via Ollama. Explanations can be requested in multiple languages (specified via the \texttt{--language} parameter), leveraging the multilingual capabilities of the chosen LLM. The generated text, along with the model identifier, is stored in the Violation KG upon a cache miss.

\paragraph{Workflow Summary}
Upon detecting a violation, xpSHACL extracts detailed violation data, builds a justification tree, generates a violation signature, queries the Violation KG, retrieves context, potentially calls an LLM for explanation generation (managing rate limits if applicable), updates the KG, and finally returns the explanation output alongside the structured violation data.
    
\section{Evaluation}\label{section:evaluation}
This section presents an empirical evaluation of the xpSHACL system, focusing primarily on its performance, efficiency gains from the Violation KG cache, and qualitative observations regarding the generated explanations. A formal user study to quantitatively assess explanation quality, user satisfaction, and helpfulness in resolving violations is planned as crucial future work.

\subsection{Research Questions addressed}
This evaluation provides initial insights related to the following research questions, with a full quantitative assessment of RQ1 and RQ4 deferred to future user studies:

\textbf{RQ1: Explanation Quality:} How clear, complete, correct, and helpful are the natural language explanations generated by xpSHACL for SHACL constraint violations?

\textbf{RQ2: Efficiency:}  What is the performance overhead of xpSHACL in terms of explanation generation time compared to standard SHACL validation, and how effectively does the Violation KG improve this efficiency over time?

\textbf{RQ3: Consistency:} Does the Violation KG ensure consistency in the explanations provided for recurring SHACL constraint violations? 

\textbf{RQ4: User Satisfaction:} How satisfied are users with the explanations and correction suggestions provided by xpSHACL, and how effectively do these explanations enable them to understand and resolve SHACL constraint violations?

\subsection{Experimental Setup}
To address the research questions, a series of experiments were conducted using public and synthetic datasets.

\subsubsection{Evaluation with Public Datasets}
\label{sec:eval_public}

To assess the applicability and performance of xpSHACL in a real-world scenario, we conducted an evaluation using publicly available ontologies from the Linked Open Vocabularies (LOV) repository and a predefined set of SHACL constraints focused on ontology quality (defined in \path{data/shark_shapes.ttl}).

\paragraph{Experimental Setup} We utilized the Linked Open Vocabularies (LOV) repository as the source for diverse RDF datasets. Using the LOV API\footnote{Available at \url{http://lov.okfn.org/dataset/lov/api/v2/vocabulary/list}, accessed on 15-04-2025}, we retrieved the URIs of \texttt{868} registered vocabularies. For the SHACL constraints, we employed the shapes defined in \texttt{data/shark\_shapes.ttl}, which encode common ontology design guidelines (e.g., requirements for labels, comments, cardinality restrictions, naming conventions).

\paragraph{Procedure}  We programmatically downloaded and parsed each ontology from LOV (\path{data/shark_tests.py}). For each successfully processed ontology, we executed the xpSHACL validation pipeline, using the ontology as the data graph and \path{data/shark\_shapes.ttl} as the shapes graph. We configured xpSHACL to use the \texttt{gemini-2.0-flash-lite} model for explanation generation and requested explanations in English (\texttt{en}). The validation process included \texttt{none} inference, as specified. Execution time for the xpSHACL process (including core validation, violation extraction, explanation generation, and Violation KG interaction) was recorded for each ontology. Results are pictured below: 

\begin{itemize}
    \item \textbf{Robustness:} From the \texttt{868} attempted LOV ontologies, xpSHACL successfully parsed and initiated validation for \texttt{431} (approx. \texttt{49.7\%}). Failures were primarily due to download errors (\texttt{304}) and parsing issues (\texttt{131}), reflecting the challenges of processing diverse real-world linked data sources. \texttt{2} graphs were parsed but found empty and skipped.

    \item \textbf{Performance:} Over the \texttt{431} successfully processed ontologies, the average total execution time per ontology for xpSHACL was \texttt{13.03} seconds. The core validation and violation extraction phase averaged \texttt{11.38} seconds. Explanation generation and retrieval averaged only \texttt{0.12} seconds for the \texttt{416} ontologies containing violations. 

    \item \textbf{Cache Effectiveness:} The Violation KG demonstrated high effectiveness. Across \texttt{2301} total explanation lookups during the run, \texttt{2289} were cache hits, resulting in an overall cache hit rate of \texttt{99.48\%}. This extremely high rate significantly minimized calls to the LLM API, contributing to the low average explanation time, as only \texttt{12} unique violation signatures required generation via the LLM for this dataset and shapes combination.

    \item \textbf{Violation Analysis:} The evaluation identified \texttt{145,910} violations across the \texttt{416} non-conformant ontologies (average \texttt{350.75} violations per non-conformant ontology), highlighting common deviations from the ontology guidelines defined in \path{shark_shapes.ttl}. The most frequent violation types encountered were Cardinality constraints (\texttt{sh:minCount}, \texttt{sh:maxCount} - \texttt{78,022} instances), SPARQL-based checks from \path{shark_shapes.ttl} (\texttt{54,096} instances), and Value Type constraints (\path{sh:datatype}, \path{sh:class}, \path{sh:nodeKind} - \texttt{13,792} instances).

    \item \textbf{Qualitative Observation of Explanation Quality:} Manual inspection of a sample of generated explanations (retrieved from \path{data/lov_explanation_outputs/} after running the \path{data/shark_tests.py} script) for violations found in ontologies like rdfs.org and xlmns.com indicated that xpSHACL  provided clear and accurate justifications based on the violation signature, with consistent cache hits. For instance, for a \texttt{sh:minCount 1} violation on \texttt{rdfs:label} for class \texttt{[\textit{ObservableProperty}]}, xpSHACL generated the following explanation: \textit{``The class `ObservableProperty` is missing a label. This means the data doesn't provide any information about a label associated with this class. The shape definition requires that this class has at least one label, but it doesn't.''}. The corresponding correction suggestions were: 
    ``
    \begin{enumerate}
        \item \textit{\textbf{Add a label:} Ensure that the class in question has a label assigned to it. This typically involves using a property like `rdfs:label` to provide a human-readable name for the class.  The label should be provided in a language such as `en` to comply with your requirements.
        \item \textbf{Use a language tag:} When adding a label, make sure to use a language tag (e.g., `@en` for English) to specify the language of the label, which is important for internationalization
        \item \textbf{Verify label existence:} Double-check that the SHACL shape defining the rule that requires labels actually exists. This can be caused by the rule not having being properly defined in the shape definition itself.
        \item \textbf{Check your SHACL shape definition:} Review the SHACL shape definition to ensure it includes a constraint that requires the presence of a label property (e.g., `sh:property` with `sh:path` to `rdfs:label`). Consider if any conditions are defined that might exempt classes from having labels in certain scenarios; this should be explicitly defined in the SHACL rules if intended.}
    \end{enumerate}
      ''
\end{itemize}

These observations suggest the LLM, guided by the justification tree and context, can produce relevant and actionable explanations. A formal, large-scale human evaluation is needed for a quantitative assessment (see Section \ref{section:conclusion}).

This evaluation demonstrates xpSHACL's ability to handle diverse, real-world RDF data and provide explainable feedback on quality issues defined by SHACL shapes. The Violation KG proves highly effective in optimizing performance, especially when applying consistent quality checks across multiple datasets, achieving a cache hit rate nearing 99.5\% in this experiment.

\subsubsection{Baseline System}
To assess the computational overhead introduced by the explainability features, we compared the execution time of xpSHACL against the baseline \texttt{pyshacl} validator. Both tools processed the same synthetic RDF dataset (2MB) and SHACL shapes file over ten consecutive runs (namely, \texttt{complex\_data.ttl} and \texttt{complex\_shapes.ttl} files inside the \texttt{data} folder of the repository). \footnote{For reproducibility, this evaluation process is described in the README file of the repository, section "Performance evaluation"}

\begin{figure*}[hpbt!]
  \centering
\includegraphics[scale=0.5,keepaspectratio]{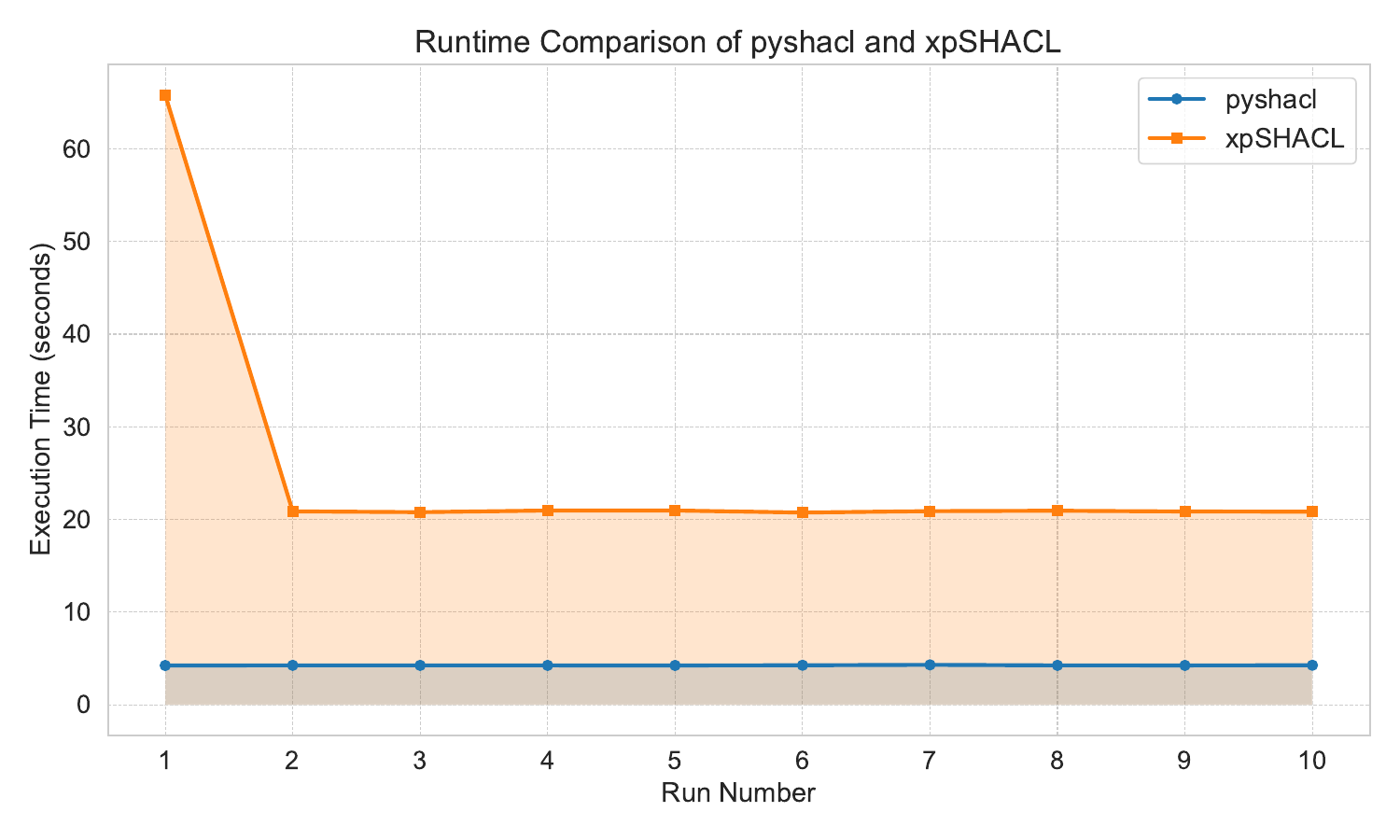}
  \caption{Runtime comparison between \texttt{pyshacl} and xpSHACL over 10 runs on the same dataset.}
  \label{fig:comp}
\end{figure*}



As illustrated in Figure \ref{fig:comp}, \texttt{pyshacl} exhibits consistent, low execution times (approx. 4 seconds per run). In contrast, xpSHACL's initial run incurs significant overhead (approx. 65 seconds). This latency is primarily attributed to the first-time processing of unique violation signatures encountered, which involves generating justification trees, retrieving domain context, making necessary calls to the configured Language Model (LLM) API for natural language explanations, and populating the Violation KG.

Subsequent runs demonstrate the effectiveness of the KG caching mechanism. By retrieving previously generated explanations for recurring violation signatures directly from the KG, xpSHACL avoids repeated LLM API calls, drastically reducing runtime to a stable level (around 20 seconds). This highlights the benefit of the KG for amortizing the cost of explanation generation over time, especially when validating evolving data against stable shapes.

However, a persistent overhead of roughly 20 seconds remains compared to \texttt{pyshacl}, even during these cached runs. This residual cost stems from xpSHACL's core explainability logic, which still executes once per unique violation signature: loading the KG, looking up signatures, constructing justification trees, and retrieving domain context, which inherently involve graph querying operations beyond the basic validation checks performed by \texttt{pyshacl}.

While the current caching significantly improves performance after the initial run, future work will focus on further minimizing the runtime overhead, particularly for these subsequent, cached executions. Promising directions include:

\begin{itemize}
    \item \textbf{Hybrid Query Engines:} Investigating the integration of high-performance RDF query engines (e.g., QLever) as specialized backends for query-intensive tasks like context retrieval, potentially offering faster execution than the underlying RDFLib query engine for specific query patterns, while retaining RDFLib for core graph management and validation compatibility.
    \item \textbf{Extended Caching:} Exploring extensions to the Violation KG to cache not only the final explanation text but also intermediate artifacts, such as the generated justification trees and retrieved domain context, further reducing redundant computations on subsequent runs.
\end{itemize}

These enhancements would reduce the gap between cached xpSHACL runs and baseline \texttt{pyshacl} validation, making explainable SHACL validation even more practical for larger datasets and interactive scenarios.

\subsubsection{Evaluation Metrics and Future Work Considerations}
Based on the empirical evaluation conducted, the following metrics were used:

\begin{itemize}
    \item \textbf{Efficiency:} Measured by execution time (total runtime, validation time, explanation generation/retrieval time) and KG cache hit rate. 
    \item \textbf{Robustness:} Measured by the success rate in processing real-world datasets (LOV ontologies).
    \item \textbf{Qualitative Explanation Quality:} Assessed through manual inspection of generated explanations for clarity, correctness, and relevance.
\end{itemize}

To fully address the research questions and provide a comprehensive evaluation of xpSHACL, particularly regarding explanation quality and user satisfaction, the following are planned as future work:

\begin{outline}
    \1 \textbf{Formal User Study (Addressing RQ1 and RQ4):} Conduct a controlled user study with participants of varying technical backgrounds (domain experts and non-technical users). This study will involve:
    \2 Presenting users with SHACL violations and the corresponding xpSHACL explanations and suggestions.
    \2 Administering questionnaires using Likert scales and open-ended questions to quantitatively and qualitatively assess clarity, completeness, correctness, helpfulness, and overall satisfaction.
    \2 Task-based evaluation where users attempt to understand and resolve violations based on the provided explanations, measuring success rate and time taken.
    \1 \textbf{Quantitative Consistency Assessment (Addressing RQ3):} Implement and apply text similarity metrics (e.g., cosine similarity of sentence embeddings, ROUGE, BLEU) to compare explanations generated for violations with the same signature over different runs or datasets, providing a quantitative measure of KG cache consistency.
    \1 \textbf{Scalability Evaluation (Further Addressing RQ2):} Conduct more extensive performance testing on larger and more complex synthetic and real-world datasets to rigorously assess the scalability of xpSHACL, particularly the impact of graph size on context retrieval and KG lookup.
\end{outline}

\textbf{Consistency (RQ3):}

\begin{itemize}
    \item For violations that share the same signature (i.e., involve the same shape, constraint, and property path), the explanations retrieved from the Violation KG for these recurring violations will be compared using text similarity metrics. Techniques such as calculating the cosine similarity of sentence embeddings will be used to quantify the degree of consistency in the explanations provided. 
\end{itemize}

\textbf{User Satisfaction (RQ4):}

\begin{itemize}
    \item After participating in the user studies, participants will be asked to complete questionnaires to gather feedback on their overall satisfaction with the explanations provided by xpSHACL, the ease of understanding, and their perceived helpfulness in resolving the violations. Qualitative feedback will also be collected through open-ended questions in the questionnaires or through follow-up interviews to gain deeper insights into the users' experiences and perceptions of xpSHACL.
\end{itemize}

\subsection{Threats to Validity}
Several potential threats to the validity of the evaluation are acknowledged. The subjectivity inherent in the qualitative assessment of explanation quality is a primary concern, which the planned formal user study with structured questionnaires and multiple evaluators will mitigate. The selection of datasets and SHACL shapes, while diverse, may not fully represent all possible SHACL use cases, a limitation common in such evaluations. The performance and characteristics of the specific LLM used (Gemini 2.0 Flash Lite) influence the quality of explanations; future work could explore the impact of using different LLMs. Finally, the generalizability of the performance results may be influenced by the specific hardware and environment used for testing. The lack of a formal user study in this initial evaluation is a significant limitation for fully addressing RQ1 and RQ4, and its completion is a high priority for future work.

\section{Conclusion and Future Work}\label{section:conclusion}
This paper introduces xpSHACL, a novel system designed to provide understandable explanations for SHACL constraint violations in RDF data in multiple languages. With a novel approach combining rule-based justification trees, retrieval-augmented generation (RAG), and LLMs, xpSHACL aims to bridge the gap between the technical output of SHACL validation engines and the comprehension needs of non-technical users. The system's architecture includes an Extended SHACL Validator, a Justification Tree Builder, a Context Retriever, an Explanation Generator, and a Violation KG, each playing a crucial role in the explanation generation process. An evaluation plan is outlined to assess the effectiveness of xpSHACL in terms of explanation quality, efficiency, consistency, and user satisfaction.

Future work may focus on the full implementation and evaluation of the xpSHACL system as described in the evaluation plan. This includes developing and refining the architectural components, conducting user studies and expert assessments, evaluating consistency and scalability, and analysing the results to assess the system's performance and effectiveness. Therefore, several research directions were identified:
\begin{itemize}
    \item \textbf{Enhancing Context Retrieval:}  Exploring more advanced techniques for retrieving relevant context, such as using KG embedding models to identify semantically similar shapes or ontology concepts, or incorporating user feedback to refine the retrieval process.
    \item \textbf{Improving Explanation Generation:}  Investigating different prompting strategies and fine-tuning techniques for the LLM to generate even more business-oriented, accurate, nuanced, and user-tailored explanations.
    \item \textbf{Collaborative violations KG:} The more we run the application, the more complete the Violation KG gets. Inspired by QuitStore \cite{arndt2016distributed}, running it collaboratively with the support of multiple users and hosting the KG in a version-control system such as Git, at some point most violations will already exist in the KG, virtually dismissing LLMs usage, enabling users to have more detailed suggestions without the need to have access to LLM API keys.
    \item \textbf{Integration with SHACL Editors and Tools:} Integrate xpSHACL capabilities with existing SHACL editors and validation tools like SHARK \cite{publio2018shark}, RDFUnit \cite{kontokostas2014test}, or even libraries like \texttt{pySHACL} and others to provide seamless and real-time explanations within the user's workflow.
    \item \textbf{User Feedback and Correction Mechanisms:} Implementing a user interface that allows users to review, verify, and edit generated explanations, thereby mitigating the propagation of inaccuracies stemming from the non-deterministic nature of LLMs and ensuring the quality and accuracy of explanations stored in the Violation KG.
    \item \textbf{Managing KG Scalability and Redundancy:} Future work will also explore strategies for managing the scalability and retrieval efficiency of the Violation KG as it grows, including methods for clustering similar violations and consolidating their explanations to avoid redundancy and optimize performance.
\end{itemize}

The development of xpSHACL represents a significant step towards making SHACL validation more accessible and actionable for a broader audience. This will contribute to improved data quality and interoperability in KG applications.

\bibliographystyle{ACM-Reference-Format}
\bibliography{custom}


\begin{thebibliography}{39}


\ifx \showCODEN    \undefined \def \showCODEN     #1{\unskip}     \fi
\ifx \showDOI      \undefined \def \showDOI       #1{#1}\fi
\ifx \showISBNx    \undefined \def \showISBNx     #1{\unskip}     \fi
\ifx \showISBNxiii \undefined \def \showISBNxiii  #1{\unskip}     \fi
\ifx \showISSN     \undefined \def \showISSN      #1{\unskip}     \fi
\ifx \showLCCN     \undefined \def \showLCCN      #1{\unskip}     \fi
\ifx \shownote     \undefined \def \shownote      #1{#1}          \fi
\ifx \showarticletitle \undefined \def \showarticletitle #1{#1}   \fi
\ifx \showURL      \undefined \def \showURL       {\relax}        \fi
\providecommand\bibfield[2]{#2}
\providecommand\bibinfo[2]{#2}
\providecommand\natexlab[1]{#1}
\providecommand\showeprint[2][]{arXiv:#2}

\bibitem[\protect\citeauthoryear{Ahmetaj, David, Ortiz, Polleres, Shehu, and Simkus}{Ahmetaj et~al\mbox{.}}{2021}]%
        {ahmetaj2021reasoning}
\bibfield{author}{\bibinfo{person}{Shqiponja Ahmetaj}, \bibinfo{person}{Robert David}, \bibinfo{person}{Magdalena Ortiz}, \bibinfo{person}{Axel Polleres}, \bibinfo{person}{Bojken Shehu}, {and} \bibinfo{person}{Mantas Simkus}.} \bibinfo{year}{2021}\natexlab{}.
\newblock \showarticletitle{{Reasoning about explanations for non-validation in SHACL}}. In \bibinfo{booktitle}{\emph{Description Logics}}.
\newblock


\bibitem[\protect\citeauthoryear{Ahmetaj, Merkl, and Pichler}{Ahmetaj et~al\mbox{.}}{2024}]%
        {ahmetaj2024consistent}
\bibfield{author}{\bibinfo{person}{Shqiponja Ahmetaj}, \bibinfo{person}{Timo~Camillo Merkl}, {and} \bibinfo{person}{Reinhard Pichler}.} \bibinfo{year}{2024}\natexlab{}.
\newblock \showarticletitle{{Consistent query answering over SHACL constraints}}.
\newblock \bibinfo{journal}{\emph{arXiv preprint arXiv:2406.16653}} (\bibinfo{year}{2024}).
\newblock


\bibitem[\protect\citeauthoryear{Arndt, Radtke, and Martin}{Arndt et~al\mbox{.}}{2016}]%
        {arndt2016distributed}
\bibfield{author}{\bibinfo{person}{Natanael Arndt}, \bibinfo{person}{Norman Radtke}, {and} \bibinfo{person}{Michael Martin}.} \bibinfo{year}{2016}\natexlab{}.
\newblock \showarticletitle{Distributed collaboration on rdf datasets using git: Towards the quit store}. In \bibinfo{booktitle}{\emph{Proceedings of the 12th International Conference on Semantic Systems}}. \bibinfo{pages}{25--32}.
\newblock


\bibitem[\protect\citeauthoryear{Berners-Lee, Connolly, Kagal, Scharf, and Hendler}{Berners-Lee et~al\mbox{.}}{2008}]%
        {berners2008n3logic}
\bibfield{author}{\bibinfo{person}{Tim Berners-Lee}, \bibinfo{person}{Dan Connolly}, \bibinfo{person}{Lalana Kagal}, \bibinfo{person}{Yosi Scharf}, {and} \bibinfo{person}{Jim Hendler}.} \bibinfo{year}{2008}\natexlab{}.
\newblock \showarticletitle{{N3Logic: A logical framework for the World Wide Web}}.
\newblock \bibinfo{journal}{\emph{Theory and Practice of Logic Programming}} \bibinfo{volume}{8}, \bibinfo{number}{3} (\bibinfo{year}{2008}), \bibinfo{pages}{249--269}.
\newblock


\bibitem[\protect\citeauthoryear{Bianchi, Rossiello, Costabello, Palmonari, and Minervini}{Bianchi et~al\mbox{.}}{2020}]%
        {bianchi2020knowledge}
\bibfield{author}{\bibinfo{person}{Federico Bianchi}, \bibinfo{person}{Gaetano Rossiello}, \bibinfo{person}{Luca Costabello}, \bibinfo{person}{Matteo Palmonari}, {and} \bibinfo{person}{Pasquale Minervini}.} \bibinfo{year}{2020}\natexlab{}.
\newblock \showarticletitle{{Knowledge graph embeddings and explainable AI}}.
\newblock In \bibinfo{booktitle}{\emph{Knowledge Graphs for Explainable Artificial Intelligence: Foundations, Applications and Challenges}}. \bibinfo{publisher}{IOS Press}, \bibinfo{pages}{49--72}.
\newblock


\bibitem[\protect\citeauthoryear{Bratanic}{Bratanic}{2024}]%
        {bratanic2024using}
\bibfield{author}{\bibinfo{person}{Toma{\v{z}} Bratanic}.} \bibinfo{year}{2024}\natexlab{}.
\newblock \bibinfo{booktitle}{\emph{Using a knowledge graph to implement a rag application}}.
\newblock \bibinfo{type}{{T}echnical {R}eport}.
\newblock
\newblock
\shownote{Accessed: 2025-03-17.}


\bibitem[\protect\citeauthoryear{Chen, Bao, Zheng, Qi, Wei, and Hu}{Chen et~al\mbox{.}}{2024}]%
        {chen2024optimizing}
\bibfield{author}{\bibinfo{person}{Jiajing Chen}, \bibinfo{person}{Runyuan Bao}, \bibinfo{person}{Hongye Zheng}, \bibinfo{person}{Zhen Qi}, \bibinfo{person}{Jianjun Wei}, {and} \bibinfo{person}{Jiacheng Hu}.} \bibinfo{year}{2024}\natexlab{}.
\newblock \showarticletitle{Optimizing Retrieval-Augmented Generation with Elasticsearch for Enhanced Question-Answering Systems}.
\newblock \bibinfo{journal}{\emph{arXiv preprint arXiv:2410.14167}} (\bibinfo{year}{2024}).
\newblock


\bibitem[\protect\citeauthoryear{Chen}{Chen}{2025}]%
        {chen2025retrieval}
\bibfield{author}{\bibinfo{person}{Ruixi Chen}.} \bibinfo{year}{2025}\natexlab{}.
\newblock \showarticletitle{Retrieval-Augmented Generation with Knowledge Graphs: A Survey}. In \bibinfo{booktitle}{\emph{Computer Science Undergradaute Conference 2025@ XJTU}}.
\newblock


\bibitem[\protect\citeauthoryear{Corman, Florenzano, Reutter, and Savkovi{\'c}}{Corman et~al\mbox{.}}{2019}]%
        {corman2019validating}
\bibfield{author}{\bibinfo{person}{Julien Corman}, \bibinfo{person}{Fernando Florenzano}, \bibinfo{person}{Juan~L Reutter}, {and} \bibinfo{person}{Ognjen Savkovi{\'c}}.} \bibinfo{year}{2019}\natexlab{}.
\newblock \showarticletitle{{Validating SHACL constraints over a SPARQL endpoint}}. In \bibinfo{booktitle}{\emph{The Semantic Web--ISWC 2019: 18th International Semantic Web Conference, Auckland, New Zealand, October 26--30, 2019, Proceedings, Part I 18}}. \bibinfo{publisher}{Springer}, \bibinfo{pages}{145--163}.
\newblock


\bibitem[\protect\citeauthoryear{da~Silva, Kocher, Gehlhoff, and Fay}{da~Silva et~al\mbox{.}}{2024}]%
        {dasilva2024use}
\bibfield{author}{\bibinfo{person}{Luis Miguel~Vieira da Silva}, \bibinfo{person}{Aljosha Kocher}, \bibinfo{person}{Felix Gehlhoff}, {and} \bibinfo{person}{Alexander Fay}.} \bibinfo{year}{2024}\natexlab{}.
\newblock \showarticletitle{On the use of large language models to generate capability ontologies}. In \bibinfo{booktitle}{\emph{2024 IEEE 29th International Conference on Emerging Technologies and Factory Automation (ETFA)}}. IEEE, \bibinfo{pages}{1--8}.
\newblock


\bibitem[\protect\citeauthoryear{Delva, Dimou, Jakubowski, and Van~den Bussche}{Delva et~al\mbox{.}}{2023}]%
        {delva2023data}
\bibfield{author}{\bibinfo{person}{Thomas Delva}, \bibinfo{person}{Anastasia Dimou}, \bibinfo{person}{Maxime Jakubowski}, {and} \bibinfo{person}{Jan Van~den Bussche}.} \bibinfo{year}{2023}\natexlab{}.
\newblock \showarticletitle{Data provenance for SHACL}. In \bibinfo{booktitle}{\emph{Proceedings 26th International Conference on Extending Database Technology (EDBT 2023)}}. OpenProceedings. org.
\newblock


\bibitem[\protect\citeauthoryear{Delva, Smedt, Oo, Assche, Lieber, and Dimou}{Delva et~al\mbox{.}}{2021}]%
        {delva2021rml2shacl}
\bibfield{author}{\bibinfo{person}{Thomas Delva}, \bibinfo{person}{Birte~De Smedt}, \bibinfo{person}{Sitt~Min Oo}, \bibinfo{person}{Dylan~Van Assche}, \bibinfo{person}{Sven Lieber}, {and} \bibinfo{person}{Anastasia Dimou}.} \bibinfo{year}{2021}\natexlab{}.
\newblock \showarticletitle{{RML2SHACL: RDF generation taking shape}}. In \bibinfo{booktitle}{\emph{Proceedings of the 11th Knowledge Capture Conference}}. \bibinfo{pages}{153--160}.
\newblock


\bibitem[\protect\citeauthoryear{Denaux, Dolbear, Hart, Dimitrova, and Cohn}{Denaux et~al\mbox{.}}{2011}]%
        {denaux2011supporting}
\bibfield{author}{\bibinfo{person}{Ronald Denaux}, \bibinfo{person}{Catherine Dolbear}, \bibinfo{person}{Glen Hart}, \bibinfo{person}{Vania Dimitrova}, {and} \bibinfo{person}{Anthony~G. Cohn}.} \bibinfo{year}{2011}\natexlab{}.
\newblock \showarticletitle{Supporting domain experts to construct conceptual ontologies: A holistic approach}.
\newblock \bibinfo{journal}{\emph{Journal of Web semantics}} \bibinfo{volume}{9}, \bibinfo{number}{2} (\bibinfo{year}{2011}), \bibinfo{pages}{113--127}.
\newblock


\bibitem[\protect\citeauthoryear{Donkers and Petrova}{Donkers and Petrova}{2024}]%
        {donkers2024converting}
\bibfield{author}{\bibinfo{person}{Alex~JA Donkers} {and} \bibinfo{person}{Ekaterina Petrova}.} \bibinfo{year}{2024}\natexlab{}.
\newblock \showarticletitle{{Converting Fire Safety Regulations to SHACL Shapes Using Natural Language Processing}}. In \bibinfo{booktitle}{\emph{Proceedings of the 3rd NLP4KGC: Natural Language Processing for Knowledge Graph Construction co-located with the 20th International Conference on Semantic Systems (SEMANTiCS 2024)}}. CEUR-WS. org.
\newblock


\bibitem[\protect\citeauthoryear{Ebrahimi, Sarker, Bianchi, Xie, Doran, and Hitzler}{Ebrahimi et~al\mbox{.}}{2018}]%
        {ebrahimi2018reasoning}
\bibfield{author}{\bibinfo{person}{Monireh Ebrahimi}, \bibinfo{person}{Md~Kamruzzaman Sarker}, \bibinfo{person}{Federico Bianchi}, \bibinfo{person}{Ning Xie}, \bibinfo{person}{Derek Doran}, {and} \bibinfo{person}{Pascal Hitzler}.} \bibinfo{year}{2018}\natexlab{}.
\newblock \showarticletitle{{Reasoning over RDF knowledge bases using deep learning}}.
\newblock \bibinfo{journal}{\emph{arXiv preprint arXiv:1811.04132}} (\bibinfo{year}{2018}).
\newblock


\bibitem[\protect\citeauthoryear{Futia and Vetr{\`o}}{Futia and Vetr{\`o}}{2020}]%
        {futia2020integration}
\bibfield{author}{\bibinfo{person}{Giuseppe Futia} {and} \bibinfo{person}{Antonio Vetr{\`o}}.} \bibinfo{year}{2020}\natexlab{}.
\newblock \showarticletitle{{On the integration of knowledge graphs into deep learning models for a more comprehensible AI—Three challenges for future research}}.
\newblock \bibinfo{journal}{\emph{Information}} \bibinfo{volume}{11}, \bibinfo{number}{2} (\bibinfo{year}{2020}), \bibinfo{pages}{122}.
\newblock


\bibitem[\protect\citeauthoryear{Gayo, Prud'Hommeaux, Boneva, and Kontokostas}{Gayo et~al\mbox{.}}{2017}]%
        {gayo2017validating}
\bibfield{author}{\bibinfo{person}{Jose Emilio~Labra Gayo}, \bibinfo{person}{Eric Prud'Hommeaux}, \bibinfo{person}{Iovka Boneva}, {and} \bibinfo{person}{Dimitris Kontokostas}.} \bibinfo{year}{2017}\natexlab{}.
\newblock \bibinfo{booktitle}{\emph{Validating {RDF} data}}.
\newblock \bibinfo{publisher}{Morgan \& Claypool Publishers}.
\newblock


\bibitem[\protect\citeauthoryear{Gercke}{Gercke}{2022}]%
        {gercke2022supporting}
\bibfield{author}{\bibinfo{person}{Julian~Alexander Gercke}.} \bibinfo{year}{2022}\natexlab{}.
\newblock \emph{\bibinfo{title}{Supporting Explainable {AI} on Semantic Constraint Validation}}.
\newblock \bibinfo{thesistype}{Master's\ thesis}. \bibinfo{school}{Hannover: Gottfried Wilhelm Leibniz Universit{\"a}t}.
\newblock


\bibitem[\protect\citeauthoryear{Herbold, Sadeghi, and Vogelsang}{Herbold et~al\mbox{.}}{2024}]%
        {herbold2024generating}
\bibfield{author}{\bibinfo{person}{Lars Herbold}, \bibinfo{person}{Mersedeh Sadeghi}, {and} \bibinfo{person}{Andreas Vogelsang}.} \bibinfo{year}{2024}\natexlab{}.
\newblock \showarticletitle{Generating context-aware contrastive explanations in rule-based systems}. In \bibinfo{booktitle}{\emph{Proceedings of the 2024 Workshop on Explainability Engineering}}. \bibinfo{pages}{8--14}.
\newblock


\bibitem[\protect\citeauthoryear{Kontokostas}{Kontokostas}{2014}]%
        {kontokostas2014test}
\bibfield{author}{\bibinfo{person}{Dimitris Kontokostas}.} \bibinfo{year}{2014}\natexlab{}.
\newblock \showarticletitle{Test-driven evaluation of linked data quality}. In \bibinfo{booktitle}{\emph{Proceedings of the 23rd international conference on World Wide Web}}. \bibinfo{pages}{747--758}.
\newblock


\bibitem[\protect\citeauthoryear{L{\'e}cu{\'e}, L{\'e}cu{\'e}, and Hitzler}{L{\'e}cu{\'e} et~al\mbox{.}}{2020}]%
        {tiddi2020knowledge}
\bibfield{author}{\bibinfo{person}{Ilaria L{\'e}cu{\'e}}, \bibinfo{person}{Freddy L{\'e}cu{\'e}}, {and} \bibinfo{person}{Pascal Hitzler}.} \bibinfo{year}{2020}\natexlab{}.
\newblock \showarticletitle{Knowledge graphs for explainable artificial intelligence: Foundations, applications and challenges}.
\newblock  (\bibinfo{year}{2020}).
\newblock


\bibitem[\protect\citeauthoryear{Li, Song, Zhou, Tian, Wang, Yang, and Zhang}{Li et~al\mbox{.}}{2024}]%
        {li2024framework}
\bibfield{author}{\bibinfo{person}{Yading Li}, \bibinfo{person}{Dandan Song}, \bibinfo{person}{Changzhi Zhou}, \bibinfo{person}{Yuhang Tian}, \bibinfo{person}{Hao Wang}, \bibinfo{person}{Ziyi Yang}, {and} \bibinfo{person}{Shuhao Zhang}.} \bibinfo{year}{2024}\natexlab{}.
\newblock \showarticletitle{A Framework of Knowledge Graph-Enhanced Large Language Model Based on Question Decomposition and Atomic Retrieval}. In \bibinfo{booktitle}{\emph{Findings of the Association for Computational Linguistics: EMNLP 2024}}. \bibinfo{pages}{11472--11485}.
\newblock


\bibitem[\protect\citeauthoryear{Luo, Li, Haffari, and Pan}{Luo et~al\mbox{.}}{2023}]%
        {luo2023reasoning}
\bibfield{author}{\bibinfo{person}{Linhao Luo}, \bibinfo{person}{Yuan-Fang Li}, \bibinfo{person}{Gholamreza Haffari}, {and} \bibinfo{person}{Shirui Pan}.} \bibinfo{year}{2023}\natexlab{}.
\newblock \showarticletitle{Reasoning on graphs: Faithful and interpretable large language model reasoning}.
\newblock \bibinfo{journal}{\emph{arXiv preprint arXiv:2310.01061}} (\bibinfo{year}{2023}).
\newblock


\bibitem[\protect\citeauthoryear{Matsumoto, Moran, Choi, Hernandez, Venkatesan, Wang, and Moore}{Matsumoto et~al\mbox{.}}{2024}]%
        {matsumoto2024kragen}
\bibfield{author}{\bibinfo{person}{Nicholas Matsumoto}, \bibinfo{person}{Jay Moran}, \bibinfo{person}{Hyunjun Choi}, \bibinfo{person}{Miguel~E Hernandez}, \bibinfo{person}{Mythreye Venkatesan}, \bibinfo{person}{Paul Wang}, {and} \bibinfo{person}{Jason~H Moore}.} \bibinfo{year}{2024}\natexlab{}.
\newblock \showarticletitle{{KRAGEN: a knowledge graph-enhanced RAG framework for biomedical problem solving using large language models}}.
\newblock \bibinfo{journal}{\emph{Bioinformatics}} \bibinfo{volume}{40}, \bibinfo{number}{6} (\bibinfo{year}{2024}), \bibinfo{pages}{btae353}.
\newblock


\bibitem[\protect\citeauthoryear{Meester, Heyvaert, Arndt, Dimou, and Verborgh}{Meester et~al\mbox{.}}{2021}]%
        {demeester2021rdf}
\bibfield{author}{\bibinfo{person}{Ben~De Meester}, \bibinfo{person}{Pieter Heyvaert}, \bibinfo{person}{D{\"o}rthe Arndt}, \bibinfo{person}{Anastasia Dimou}, {and} \bibinfo{person}{Ruben Verborgh}.} \bibinfo{year}{2021}\natexlab{}.
\newblock \showarticletitle{{RDF graph validation using rule-based reasoning}}.
\newblock \bibinfo{journal}{\emph{Semantic Web}} \bibinfo{volume}{12}, \bibinfo{number}{1} (\bibinfo{year}{2021}), \bibinfo{pages}{117--142}.
\newblock


\bibitem[\protect\citeauthoryear{Mohammed, Naumann, and Harmouch}{Mohammed et~al\mbox{.}}{2025}]%
        {mohammed2025}
\bibfield{author}{\bibinfo{person}{Sedir Mohammed}, \bibinfo{person}{Felix Naumann}, {and} \bibinfo{person}{Hazar Harmouch}.} \bibinfo{year}{2025}\natexlab{}.
\newblock \showarticletitle{Step-by-Step Data Cleaning Recommendations to Improve ML Prediction Accuracy}. In \bibinfo{booktitle}{\emph{Proceedings 28th International Conference on Extending Database Technology (EDBT) 2025, Barcelona, Spain, March 25-28}}. \bibinfo{pages}{542--554}.
\newblock
\urldef\tempurl%
\url{https://doi.org/10.48786/EDBT.2025.43}
\showDOI{\tempurl}


\bibitem[\protect\citeauthoryear{Mullins}{Mullins}{2023}]%
        {mullins2023shape}
\bibfield{author}{\bibinfo{person}{Brett Mullins}.} \bibinfo{year}{2023}\natexlab{}.
\newblock \showarticletitle{The Shape of Explanations: A Topological Account of Rule-Based Explanations in Machine Learning}.
\newblock \bibinfo{journal}{\emph{arXiv preprint arXiv:2301.09042}} (\bibinfo{year}{2023}).
\newblock


\bibitem[\protect\citeauthoryear{Okulmus and {\v{S}}imkus}{Okulmus and {\v{S}}imkus}{2024}]%
        {okulmus2024shacl}
\bibfield{author}{\bibinfo{person}{Cem Okulmus} {and} \bibinfo{person}{Mantas {\v{S}}imkus}.} \bibinfo{year}{2024}\natexlab{}.
\newblock \showarticletitle{{SHACL} Validation under the {Well-founded} Semantics}. In \bibinfo{booktitle}{\emph{Proceedings of the International Conference on Principles of Knowledge Representation and Reasoning}}, Vol.~\bibinfo{volume}{21}. \bibinfo{pages}{553--562}.
\newblock


\bibitem[\protect\citeauthoryear{Pareti and Konstantinidis}{Pareti and Konstantinidis}{2021}]%
        {pareti2021review}
\bibfield{author}{\bibinfo{person}{Paolo Pareti} {and} \bibinfo{person}{George Konstantinidis}.} \bibinfo{year}{2021}\natexlab{}.
\newblock \showarticletitle{{A review of SHACL: from data validation to schema reasoning for RDF graphs}}.
\newblock \bibinfo{journal}{\emph{Reasoning Web International Summer School}} (\bibinfo{year}{2021}), \bibinfo{pages}{115--144}.
\newblock


\bibitem[\protect\citeauthoryear{Polleres, Hogan, Delbru, and Umbrich}{Polleres et~al\mbox{.}}{2013}]%
        {polleres2013rdfs}
\bibfield{author}{\bibinfo{person}{Axel Polleres}, \bibinfo{person}{Aidan Hogan}, \bibinfo{person}{Renaud Delbru}, {and} \bibinfo{person}{J{\"u}rgen Umbrich}.} \bibinfo{year}{2013}\natexlab{}.
\newblock \showarticletitle{{RDFS and OWL reasoning for linked data}}.
\newblock In \bibinfo{booktitle}{\emph{Reasoning Web International Summer School}}. \bibinfo{publisher}{Springer}, \bibinfo{pages}{91--149}.
\newblock


\bibitem[\protect\citeauthoryear{Publio}{Publio}{2018}]%
        {publio2018shark}
\bibfield{author}{\bibinfo{person}{Gustavo~Correa Publio}.} \bibinfo{year}{2018}\natexlab{}.
\newblock \showarticletitle{{SHARK: A test-driven framework for design and evolution of ontologies}}. In \bibinfo{booktitle}{\emph{The Semantic Web: ESWC 2018 Satellite Events: ESWC 2018 Satellite Events, Heraklion, Crete, Greece, June 3-7, 2018, Revised Selected Papers 15}}. \bibinfo{pages}{314--324}.
\newblock


\bibitem[\protect\citeauthoryear{Rajabi and Etminani}{Rajabi and Etminani}{2024}]%
        {rajabi2024knowledge}
\bibfield{author}{\bibinfo{person}{Enayat Rajabi} {and} \bibinfo{person}{Kobra Etminani}.} \bibinfo{year}{2024}\natexlab{}.
\newblock \showarticletitle{Knowledge-graph-based explainable {AI}: A systematic review}.
\newblock \bibinfo{journal}{\emph{Journal of information science}} \bibinfo{volume}{50}, \bibinfo{number}{4} (\bibinfo{year}{2024}), \bibinfo{pages}{1019--1029}.
\newblock


\bibitem[\protect\citeauthoryear{RDF4J}{RDF4J}{2020}]%
        {rdf4jshacl}
\bibfield{author}{\bibinfo{person}{Eclipse RDF4J}.} \bibinfo{year}{2020}\natexlab{}.
\newblock \bibinfo{title}{Validation With {SHACL}}.
\newblock \bibinfo{howpublished}{https://rdf4j.org/documentation/programming/shacl/}.
\newblock
\newblock
\shownote{Accessed: 2025-03-16.}


\bibitem[\protect\citeauthoryear{Sommer, Car, and Yu}{Sommer et~al\mbox{.}}{2021}]%
        {sommer2021pyshacl}
\bibfield{author}{\bibinfo{person}{Ashley Sommer}, \bibinfo{person}{Nicholas Car}, {and} \bibinfo{person}{Jonathan Yu}.} \bibinfo{year}{2021}\natexlab{}.
\newblock \showarticletitle{{pySHACL}}.
\newblock \bibinfo{journal}{\emph{DOI: https://doi.org/10.5281/zenodo.4750840}} (\bibinfo{year}{2021}).
\newblock


\bibitem[\protect\citeauthoryear{Song, Wang, Li, and Lin}{Song et~al\mbox{.}}{2024}]%
        {song2024good}
\bibfield{author}{\bibinfo{person}{Yifan Song}, \bibinfo{person}{Guoyin Wang}, \bibinfo{person}{Sujian Li}, {and} \bibinfo{person}{Bill~Yuchen Lin}.} \bibinfo{year}{2024}\natexlab{}.
\newblock \showarticletitle{{The good, the bad, and the greedy: Evaluation of LLMs should not ignore non-determinism}}.
\newblock \bibinfo{journal}{\emph{arXiv preprint arXiv:2407.10457}} (\bibinfo{year}{2024}).
\newblock


\bibitem[\protect\citeauthoryear{Team}{Team}{2023}]%
        {xaqt2023mastering}
\bibfield{author}{\bibinfo{person}{XAQT Team}.} \bibinfo{year}{2023}\natexlab{}.
\newblock \bibinfo{title}{Mastering Rule-Based Systems: Implementation, Benefits, and Best Practices}.
\newblock \bibinfo{howpublished}{https://www.xaqt.com/blog/mastering-rule-based-systems/}.
\newblock
\newblock
\shownote{Accessed: 2025-03-17.}


\bibitem[\protect\citeauthoryear{W3C}{W3C}{2017}]%
        {w3cshacl}
\bibfield{author}{\bibinfo{person}{World Wide Web~Consortium W3C}.} \bibinfo{year}{2017}\natexlab{}.
\newblock \bibinfo{title}{Shapes Constraint Language ({SHACL})}.
\newblock \bibinfo{howpublished}{https://www.w3.org/TR/shacl/}.
\newblock
\newblock
\shownote{Accessed: 2025-03-17.}


\bibitem[\protect\citeauthoryear{Zhang, Ye, Liu, Ren, Wu, and Chen}{Zhang et~al\mbox{.}}{2024}]%
        {zhang2024knowledge}
\bibfield{author}{\bibinfo{person}{Mengqi Zhang}, \bibinfo{person}{Xiaotian Ye}, \bibinfo{person}{Qiang Liu}, \bibinfo{person}{Pengjie Ren}, \bibinfo{person}{Shu Wu}, {and} \bibinfo{person}{Zhumin Chen}.} \bibinfo{year}{2024}\natexlab{}.
\newblock \showarticletitle{Knowledge graph enhanced large language model editing}.
\newblock \bibinfo{journal}{\emph{arXiv preprint arXiv:2402.13593}} (\bibinfo{year}{2024}).
\newblock


\bibitem[\protect\citeauthoryear{Zhu, Guo, Cao, Li, and Gong}{Zhu et~al\mbox{.}}{2024}]%
        {zhu2024structugraphrag}
\bibfield{author}{\bibinfo{person}{Xishi Zhu}, \bibinfo{person}{Xiaoming Guo}, \bibinfo{person}{Shengting Cao}, \bibinfo{person}{Shenglin Li}, {and} \bibinfo{person}{Jiaqi Gong}.} \bibinfo{year}{2024}\natexlab{}.
\newblock \showarticletitle{{StructuGraphRAG: Structured Document-Informed Knowledge Graphs for Retrieval-Augmented Generation}}. In \bibinfo{booktitle}{\emph{Proceedings of the AAAI Symposium Series}}, Vol.~\bibinfo{volume}{4}. \bibinfo{pages}{242--251}.
\newblock


\end{thebibliography}

\end{document}